\documentstyle[12pt]{article}
\renewcommand{\section}[1]{\refstepcounter{section}
\vspace{24pt}\noindent{\bf\arabic{section}.\quad #1}
\vspace*{12pt}}
\newcommand{\ulsect}[1]{\vspace{18pt}\noindent{\bf #1}
\vspace*{12pt}}

\begin{document}
\begin{flushright} CERN-TH.7037/93\\
\end{flushright}
\vspace*{10mm}
\begin{center}
{\bf What can we learn from a second phi meson peak\\
in ultrarelativistic nuclear collisions?}\\[10mm]
Che Ming Ko$^a$ and David Seibert$^{b*}$\\[5mm]
$^a$Cyclotron Institute and Physics Department,\\
Texas A\&M University, College Station, Texas 77843\\
$^b$Theory Division, CERN, CH-1211 Geneva 23, Switzerland\\[10mm]
{\bf Abstract}\\
\end{center}
\hspace*{12pt}
The decay width of a phi meson is reduced from its vacuum value
as its mass decreases in hot hadronic matter as a result of
the partial restoration of chiral symmetry.  This reduction is,
however, cancelled by collisional broadening through
the reactions $\phi\pi\to KK^*$, $\phi K\to\phi K$, $\phi\rho\to KK$,
and $\phi\phi\to KK$.  The resulting phi meson width in hot hadronic
matter is found to be less than about 10 MeV for temperatures below
200 MeV.  If hadronic
matter has a strong first-order phase transition, this narrow phi
meson with reduced mass will appear as a second peak in the dilepton
spectrum in ultrarelativistic heavy ion collisions.  We discuss use
of this second phi peak to determine the transition temperature and
the lifetime of the two-phase coexistence region in the case of a
strong first-order phase transition.  We also discuss using the peak
to determine the range of temperatures over which the transition
occurs in the case of a smooth but fast change in the entropy
density.\\
\vfill
\begin{center}
{\em Submitted to Phys.\ Rev.\ C}
\end{center}
\vfill
CERN-TH.7037/93\\
December 1993\\
\vspace*{10mm}
\footnoterule
\vspace*{3pt}
$^*$Current address: Physics Department, Kent State University,
Kent, OH 44242.  Internet: seibert@scorpio.kent.edu.\\
\newpage\setcounter{page}{1} \pagestyle{plain}
     \setlength{\parindent}{12pt}

Recently, the study of vector meson properties in hot and dense matter has
attracted much attention [\ref{ha92}].  Asakawa and Ko have shown via the
QCD sum rules that the phi meson mass decreases in hot hadronic matter
because of the appreciable number of strange particles [\ref{as93}]. The
phi meson mass is found to drop below twice the free kaon mass when the
temperature is above about 150 MeV, as shown in Fig.~1.
This finding is supported by
preliminary results from lattice gauge calculations [\ref{bo93}].
Since the kaon mass does not change
much with temperature, the phi meson can only decay into a pion and a rho
meson. Even including the decrease of the rho meson mass at finite
temperatures, the decay width of a phi meson is substantially reduced from
its width ($\sim 4$ MeV) in free space.

Asakawa and Ko used this decrease of the phi meson mass to suggest a new
signature for the quark-gluon plasma (QGP) to hadronic matter transition in
ultrarelativistic heavy-ion collisions [\ref{ko93}]. Specifically, it was
shown that if there is a strong first-order phase transition between QGP
and hadronic matter, then a double phi peak structure appears in the
dilepton invariant mass spectrum. The low mass phi peak results from the
decay of phi mesons with reduced in-medium mass in the mixed phase.
Furthermore, they pointed out that due to the small transverse expansion of
the matter during the phase transition, the transverse momentum
distribution of these low mass phi mesons offers a viable means for
determining the transition temperature.

In the above study, the change of the phi meson width in the medium is,
however, not included. Shuryak and collaborators [\ref{ls91},\ref{sh92}]
found that if the phi meson mass is assumed to be unchanged at finite
temperatures its width is then approximately doubled as a result of the
attractive kaon potential. With the phi meson mass reduced to much below two
kaon masses
at high temperatures, this effect will not be important in our studies.
However, there will be collisional broadening of the phi meson width due
to its interaction with pions. Bi and Rafelski [\ref{bi91}] estimated
that this would also double the phi meson width. In this paper, we
investigate explicitly the collisional broadening of the phi meson width
using the in-medium mass in hot hadronic matter.  We then discuss the use
of this secondary phi peak in studying the physics of ultra-relativistic
nuclear collisions.

\section{The phi meson width}

The interaction of a phi meson in hot baryon-free hadronic matter
is mainly through the reactions
$\phi\pi\to KK^*$, $\phi K\to\phi K$, $\phi\rho\to KK$,
and $\phi\phi\to KK$
shown in Fig.~2. The interaction Lagrangians needed to evaluate these
diagrams are given by
[\ref{br91},\ref{ko91}]
\begin{eqnarray}
L_{\phi KK}&=&\,ig_{\phi KK}[\bar K(\partial_\mu K)-(\partial_\mu\bar K)
K]\phi^\mu, \\
L_{K^*K\pi}&=&\,ig_{K^*K\pi}{\bar{K^*}}^\mu
{\vec\tau}\cdot [K(\partial_\mu{\vec\pi})-(\partial_\mu K)
{\vec\pi}]+\,h.c., \\
L_{\rho KK}&=&\,ig_{\rho KK}[\bar K{\vec\tau}
(\partial_\mu K)-(\partial_\mu\bar K){\vec\tau}K]\cdot{\vec\rho}^\mu.
\end{eqnarray}
The coupling constants $g_{\phi KK}$ and $g_{K^* K\pi}$ can
be determined from the $\phi$ and $K^*$ widths, i.e.,
\begin{eqnarray}
\Gamma_\phi&=&\,\frac{g_{\phi KK}^2}{24\pi}\frac{(m_\phi^2-4m_K^2)^{3/2}}
{m_\phi^2}, \\
\Gamma_{K^*}&=&\,\frac{g_{K^* K\pi}^2}{
16\pi}\frac{\{[m_{K^*}^2-
(m_K+m_\pi)^2][m_{K^*}^2-(m_K-m_\pi)^2]\}^{3/2}}{m_{K^*}^5}.
\end{eqnarray}
Using the measured widths $\Gamma_\phi\sim 3.7$ MeV and $\Gamma_{K^*}\sim
51.3$
MeV for the decays $\phi\to \bar KK$ and ${K^*}^+\to K^+\pi^0$, we obtain
$g_{\phi KK}^2/(4\pi)\sim 1.82$ and $g_{K^* K\pi}^2/(4\pi)\sim 0.86$.

The $\rho KK$ coupling is one-half of the $\rho\pi\pi$ coupling
according to the quark model [\ref{lo90}].  From the width of the rho
meson,
\begin{equation}
\Gamma_\rho=\,\frac{g_{\rho\pi\pi}^2}{48\pi}\frac{(m_\rho^2-4m_\pi^2)^{3/2}}
{m_\rho^2}\sim 153\,{\rm MeV},
\end{equation}
one obtains $g_{\rho\pi\pi}^2/(4\pi)\sim 2.94$ which leads to
$g_{\rho KK}^2/(4\pi)\sim 0.735$.

Averaging the square of the invariant scattering matrix over initial
(and summing over final) isospins and polarizations, we obtain for the
reaction $\phi\pi\to KK^*$,
\begin{eqnarray}\label{m}
{\overline{M^2}}(\phi\pi\to KK^*)&=
&\,\frac{4g_{\phi KK}^2g_{K^*K\pi}^2}{3(t-m_K^2)^2}\,
[m_\phi^2-2m_K^2-2t+\frac{(m_K^2-t)^2}{m_\phi^2}]\nonumber\\
&&\times [m_{K^*}^2-2m_\pi^2-2t+\frac{(m_\pi^2-t)^2}{m_{K^*}^2}].
\end{eqnarray}
In the above, the four-momentum transfer $t$ is given by
\begin{equation}{\label{q2}}
t=\,m_\phi^2+m_K^2-2(m_\phi^2+p^2)^{1/2}(m_K^2+{p^\prime}^2)^{1/2}
+2pp^\prime \cos\theta,
\end{equation}
where $\theta$ is the scattering angle in the center-of-momentum (cm)
frame, and the initial and final cm momenta are given by
\begin{equation}{\label{p}}
p,p'=\,\sqrt{[s-(m_1+m_2)^2][s-(m_1-m_2)^2]/(4s)}.
\end{equation}
Here $s$ is the square of the cm energy of the two
interacting particles with masses $m_1$ and $m_2$.

The exchanged kaon in $\phi\pi\to KK^*$
can be on mass shell if the phi meson mass is greater than twice the
kaon mass.  In this case, the invariant amplitude becomes
singular.  This singularity is, however, removed by the imaginary
part of the kaon propagator in the hot matter. In Ref.~[\ref{sh92}],
the imaginary part of the kaon potential in the medium
has been evaluated and is shown to be about 10 MeV at temperature
$T=\,$150 MeV, increasing to about
30 MeV at $T=\,$200 MeV.  For simplicity, we take the width of a
kaon to be $\Gamma_K\sim\,$20 MeV and
replace the square of the kaon propagator in eq.(\ref{m}) by
$(t-m_K^2)^2+(m_K\Gamma_K)^2$.

The averaged and summed squared invariant amplitude for the reaction
$\phi K\to\phi K$ is
\begin{eqnarray}
{\overline{M^2}}(\phi K\to\phi K)&=
&\,\frac{g_{\phi KK}^4}{3(t-m_K^2)^2}\,
[m_\phi^2-2m_K^2-2t+\frac{(m_K^2-t)^2}{m_\phi^2}]^2.
\end{eqnarray}
Here, $t$ is defined as in eq.(\ref{q2}) with the
kaon mass replaced by the pion mass.
For the reactions $\phi\rho\to KK$ and $\phi\phi\to KK$,
the averaged and summed squared invariant amplitudes are given by
\begin{eqnarray}
{\overline{M^2}}(\phi\rho\to KK)&=
&\,\frac{4g_{\phi KK}^2g_{\rho KK}^2}{9(t-m_K^2)^2}\,
[m_\phi^2-2m_K^2-2t+\frac{(m_K^2-t)^2}{m_\phi^2}]\nonumber\\
&&\times [m_\rho^2-2m_K^2-2t+\frac{(m_K^2-t)^2}{m_\rho^2}], \\
{\overline{M^2}}(\phi\phi\to KK)&=
&\,\frac{4g_{\phi KK}^4}{9(t-m_K^2)^2}\,
[m_\phi^2-2m_K^2-2t+\frac{(m_K^2-t)^2}{m_\phi^2}]^2,
\end{eqnarray}
respectively.  In all cases, the four-momentum transfer $t$ is
defined in accordance with the Feynman graphs of Fig.~2, and the
initial and final cm momenta are given by eq.(\ref{p}) with the
appropriate masses.

Using Boltzmann distributions for phi mesons and pions,
the rate for the reaction $\phi \pi \to KK^{*}$ is
\begin{equation}
\Gamma=\,\frac{3T}{64 \pi^3 m_\phi^2 K_2(m_\phi/T)}
\int_{z_0}^\infty dz \, K_1(z) \, p \, p'
\int_{-1}^1 d\cos\theta {\overline{M^2}} (s,\cos\theta).
\end{equation}
In the above, $K_1$ and $K_2$ are modified Bessel functions of the first
and second kind, $s=(zT)^2$, and
$z_0=\mbox{max}\{(m_\phi+m_\pi)/T,(m_K+m_{K^*})/T\}$.
Including Bose enhancement factors
should introduce corrections of order ten percent to this result.
Similar equations can be obtained for the reactions
$\phi K\to\phi K$, $\phi\rho\to KK$, and $\phi\phi\to KK$.
The collisional width of the phi meson is then given by
\begin{equation}
\Gamma^{coll}_\phi(T)=\, \Gamma_{\phi\pi\to KK^*} +\Gamma_{\phi K\to\phi K}
+\Gamma_{\phi\rho\to KK} +\Gamma_{\phi\phi\to KK}.
\end{equation}

The width of a phi meson for $T=90-190$ MeV is shown in Fig.~3.
We show here also the temperature dependence of the phi meson
decay width into $KK$ and $\rho\pi$, assuming that
the masses of $K$, $\rho$, and $\pi$ remain unchanged at finite
temperatures.  The partial collisional widths are shown as indicated
in the figure, taking $\Gamma_K=20$ MeV for all curves.
We see that the broadening due to $\phi\pi\to KK^*$
is larger than that from $\phi K\to\phi K$, $\phi\rho\to KK$,
and $\phi\phi\to KK$.  The total width of the phi meson
is shown by the solid curve and is less than 35 MeV for $T<190$ MeV.

The phi meson can also interact elastically with a pion through the
rho meson exchange.  The $\phi\rho\pi$ coupling constant is, however,
an order of magnitude smaller than the $\phi KK$ coupling constant, as
shown by the
smaller branching ratio for phi decay into pion and rho meson. As a result,
the phi-pion elastic scattering, which is proportional to the fourth power
of the $\phi\rho\pi$ coupling, has a very small cross section.  The
phi collisional width due to this reaction turns out to be only a few KeV.
The reactions $\phi\rho\to\phi\rho$, $\phi\phi\to\pi\pi$,
and $\phi\phi\to\rho\rho$ are also proportional to the fourth power
of the $\phi\rho\pi$ coupling and are expected to be insignificant
as well.  The reactions $\phi\pi\to\rho\rho$ and $\phi K\to\phi K^*$
involve the square of the $\phi\rho\pi$ coupling, so their contributions
to the phi collisional width are also negligible.

To take into account the complicated structure of the strong interaction
vertices, we introduce at the vertex a monopole form factor,
\begin{equation}
F(t)=\,\frac{\Lambda^2-m_K^2}{\Lambda^2-t}, \label{evff}
\end{equation}
where $\Lambda$ is the cutoff parameter whose value we take from
Refs.~[\ref{holz89}] and [\ref{br91}], i.e., $\Lambda_{\phi KK}\sim
\Lambda_{K^*K\pi}\sim \Lambda_{\rho KK}\sim 1.8$ GeV.
We show the width with the form factor in Fig.~4, using $\Gamma_K
=10$ and 30 MeV; the dependence on $\Gamma_K$ is clearly unimportant
for $T=150-190$ MeV.
The total width at $T=190$ MeV is reduced by approximately a factor of
four with the form factor, so that the width is less than 10 MeV for
$T=150-190$ MeV.

\section{Extracting physical parameters}

As the secondary phi peak is narrow, there is a great possibility to
extract information about the hadronic equation of state and the dynamics
of ultra-relativistic nuclear collisions from its properties if the peak is
experimentally observed.  In this section, we discuss methods for
determining the hadronic transition temperature, $T_{\mbox{c}}$ and the
lifetime of the mixed phase, $\tau_{\mbox{m}}$.  If the transition is
not first-order, the range of temperatures over which the transition occurs,
$\delta T$, may also be measurable.

The technique for determining
$T_{\mbox{c}}$ is discussed in Ref.~[\ref{ko93}].  This is similar to a
proposal to use the transverse mass distribution of dileptons from $\rho^0$
mesons to measure $T_{\mbox{c}}$ [\ref{rrho}], but it has at least three
advantages over that proposal.  First, the second phi peak comes only from
matter in the mixed phase (at $T_{\mbox{c}}$), so there is little
contamination from phi mesons at temperatures below $T_{\mbox{c}}$.
Second, the transverse flow is still small during the mixed phase period,
giving less distortion of the transverse mass distribution from a stationary
thermal distribution.  Third, because the peak is narrower, subtractions
from background will be easier as the background changes less from one side
of the peak to the other.

The most serious of disadvantage of the phi measurement is that if there is
not a strong first-order transition, there may not be a secondary phi peak,
so the phi measurement  as proposed in Ref.~[\ref{ko93}] may be impossible.
We find this not to be a problem as long as the transition is reasonably
sharp.  For example, suppose
that the entropy density of the hot matter changes by a large factor
over a small temperature range $\delta T$, centered on some transition
temperature $T_t$.  In this case, there will be a second phi peak, but it
will be wider than a peak from fixed temperature hadronic matter. The width
of the second peak will be
\begin{equation}
\Gamma^2 ~\simeq~ \Gamma^2 (T_t) \, + \, \left( \delta T \,
\frac {d m_{\phi}} {dT} \right)_{T_t}^2 \, . \label{edelG}
\end{equation}

{}From Fig.~1, we estimate $dm_{\phi}/dT \simeq -3$ for $T=150-190$ MeV,
which is considered a likely range for $T_t$, so we would expect that
$\Gamma \simeq 3\delta T$ from the width of the transition alone.  Unless
the transition occurs over $\delta T < 3$ MeV, the width of the second phi
peak should be dominated by the contribution from $\delta T$.  If this
width is small enough, the peak will be observable; the exact width needed
depends on experimental statistics and on the dilepton background which is
not computed here.  The only hard limit for observability is that
$\Gamma~\mbox{or}~\Gamma_x < m_{\phi}(T_t)-m_x$ for all other nearby
dilepton peaks $x$, so that the individual peaks can be separated.  This
condition should be satisfied as long as $\delta T < 20-30$, given the
$T$-dependence of $m_{\phi}$ from Ref.~[\ref{ko93}].  Thus, it seems that
determination of $T_t$ is feasible as long as there is a sharp transition,
whether it is a first-order phase transition or not.

If the phi width is small throughout the collision, the second peak can
also be used to determine the lifetime of the mixed phase, $\tau_{\mbox{m}}$.
The number of phi mesons in the secondary peak is
\begin{equation}
N_s = N_{\phi} \Gamma_{\phi \rightarrow l^+l^-} \tau_{\mbox{m}}/2,
\label{eNs} \end{equation}
where $N_{\phi}$ is the number of
phi mesons that are created in the mixed phase.  [The factor of two appears
because we assume that the creation rate for phi mesons is approximately
constant, so that the average phi meson is created halfway through the mixed
phase lifetime.]  If the width is small, then not many phis will decay before
the hadronic matter freezes out, so by detailed balance there must not be
many phis created after the phase transition,
so $N_{\phi}$ is approximately equal
to the total number of phis created during the collision.  Most of these
phis will have approximately their zero-temperature mass when they decay,
so they will add to the original phi peak:
\begin{equation}
N_0 = N_{\phi} \Gamma_{\phi \rightarrow l^+l^-} / \Gamma_{\phi \rightarrow X}.
\label{eN0} \end{equation}
Combining eqs.~(\ref{eNs}) and (\ref{eN0}), we obtain
\begin{equation}
\tau_{\mbox{m}} = \frac {2 N_s} {N_0 \Gamma_{\phi \rightarrow X}}.
\end{equation}

Here we have assumed that $\Gamma_{\phi \rightarrow l^+l^-}$ and
$\Gamma_{\phi \rightarrow X}$ are constant for illustrating the technique,
but this could easily be relaxed if one wants to use more detailed models
of ultra-relativistic nuclear collisions.  In any case, we would expect
that the ratio of the two peaks will provide a good measure of
$\tau_{\mbox{m}}$ as long as $\Gamma_{\phi \rightarrow X}$ remains small.
Note that the validity of this inference depends on the width of the
phi at fixed temperature and not on the width of the second phi peak, so
the technique may be valid even if a relatively wide second phi peak is
observed as a result of a transition over a finite temperature range.

Finally, if the transition occurs over a range of temperatures $\delta T$,
it may be possible to determine $\delta T$ from eq.~(\ref{edelG}), assuming
that the phi width can be calculated accurately as a function of $T$.
This is unlikely to be feasible if $\delta T$ is small, but if the observed
width is large then it might be possible to get a reasonably accurate
estimate of $\delta T$.  For example, if the observed width is three times
the estimated thermal width then it should be possible to determine
$\delta T$ with 10\% accuracy if the thermal width can be calculated
50\% accuracy.

\section{Summary}

We have evaluated the width of a phi meson in a hot hadronic matter.
The reduction of the phi meson width due to the possible decrease of its
mass at high temperature is found to be cancelled by the
collisional broadening through the reactions $\phi\pi\to KK^*$,
$\phi K\to\phi K$, $\phi\rho\to KK$, and $\phi\phi\to KK$.
The resulting phi meson width at finite temperatures is not very much
larger than its width in free space.
The narrow phi meson width justifies the assumption of Ref.~[\ref{ko93}].
If there is a strong first-order phase transition between
the quark-gluon plasma and the hadronic matter in ultrarelativistic
heavy-ion collisions, then a low mass secondary phi peak is expected
to be observed in the dilepton spectrum.
This second peak allows us to infer the transition temperature and
the lifetime of the mixed phase in the case of a first-order transition,
and also the range of temperatures over which the transition takes place
in the case of a smooth but fast transition.

\bigskip

We thank Scott Pratt, Edward Shuryak and Chungsik Song
for helpful discussions. The work of C.M.K was
supported in part by the National Science Foundation under Grant No.\
PHY-9212209 and the Welch Foundation under Grant No.\ A-1110.  This
material is also based upon work supported by the North Atlantic Treaty
Organization under a Grant awarded in 1991.

\vfill \eject

\ulsect{References}

\begin{list}{\arabic{enumi}.\hfill}{\setlength{\topsep}{0pt}
\setlength{\partopsep}{0pt} \setlength{\itemsep}{0pt}
\setlength{\parsep}{0pt} \setlength{\leftmargin}{\labelwidth}
\setlength{\rightmargin}{0pt} \setlength{\listparindent}{0pt}
\setlength{\itemindent}{0pt} \setlength{\labelsep}{0pt}
\usecounter{enumi}}

\item \label{ha92}  T. Hatsuda, Nucl.\ Phys.\ {\bf A544}, 27c (1992).

\item \label{as93}  M. Asakawa and C. M. Ko, submitted to Nucl.\ Phys.\
{\bf A}.

\item \label{bo93}  G. Boyd, in Proc. of Quark Matter '93, ed. H. Gustafsson
{\it et al.}, Nucl.\ Phys.\ {\bf A}, in press.

\item \label{ko93}  C. M. Ko and M. Asakawa,
in Proc. of Quark Matter '93, ed. H. Gustafsson
{\it et al.}, Nucl.\ Phys.\ {\bf A}, in press.

\item \label{ls91}  D. Lissauer and E. Shuryak, Phys.\ Lett.\ B {\bf 253},
15 (1991).

\item \label{sh92}  E. Shuryak and V. Thorsson, Nucl.\ Phys.\ {\bf A536},
739 (1992).

\item \label{bi91}  P. Z. Bi and J. Rafelski, Phys.\ Lett.\ B {\bf 262},
485 (1991).

\item \label{br91}  C. M. Ko, Z. G. Wu, L. H. Xia, and G. E. Brown,
Phys.\ Rev.\ Lett.\ {\bf 66}, 2577 (1991); Phys.\ Rev.\ C {\bf 43}, 1881
(1991).

\item \label{ko91}  C. M. Ko and B. H. Sa, Phys.\ Lett.\ B {\bf 258}, 6
(1991).

\item \label{lo90} D. Lohse, J. W. Durso, K. Holinde, and J. Speth,
Phys.\ Lett.\ B {\bf 234}, 235 (1990); Nucl.\ Phys.\ {\bf A516}, 513 (1990).

\item \label{holz89} B. Holzenkamp, K. Holinde, and J. Speth, Nucl.\
Phys.\ {\bf A500}, 485 (1989).

\item \label{rrho} D. Seibert, Phys.\ Rev.\ Lett.\ {\bf 68} (1992) 1476;
D. Seibert, V.K. Mishra, and G. Fai, Phys.\ Rev.\ C {\bf 46} (1992) 330.

\end{list}

\vfill\eject

\ulsect{Figure captions}

\begin{list}{\arabic{enumi}.\hfill}{\setlength{\topsep}{0pt}
\setlength{\partopsep}{0pt} \setlength{\itemsep}{0pt}
\setlength{\parsep}{0pt} \setlength{\leftmargin}{\labelwidth}
\setlength{\rightmargin}{0pt} \setlength{\listparindent}{0pt}
\setlength{\itemindent}{0pt} \setlength{\labelsep}{0pt}
\usecounter{enumi}}

\item The phi meson mass in hot hadronic matter from QCD sum rules
[\ref{as93}].

\item Feynman diagrams for the reactions (a) $\phi\pi\to KK^*$,
(b) $\phi K\to\phi K$, (c) $\phi\rho\to KK$, and
(d) $\phi\phi\to KK$.

\item The phi meson width in hot hadronic matter.

\item The phi meson width in hot hadronic matter with vertex
form factors (\ref{evff}).

\end{list}

\vfill \eject

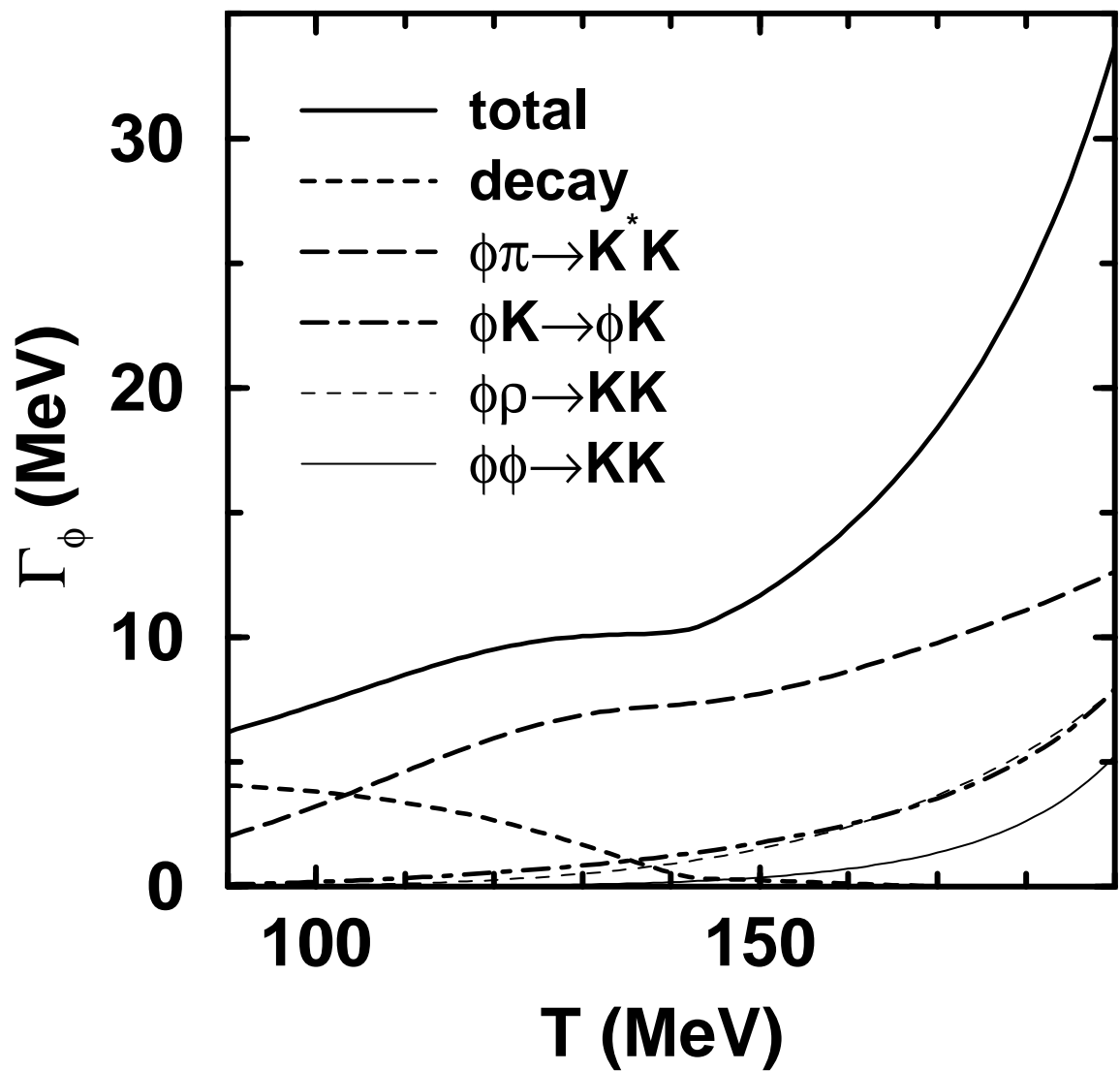
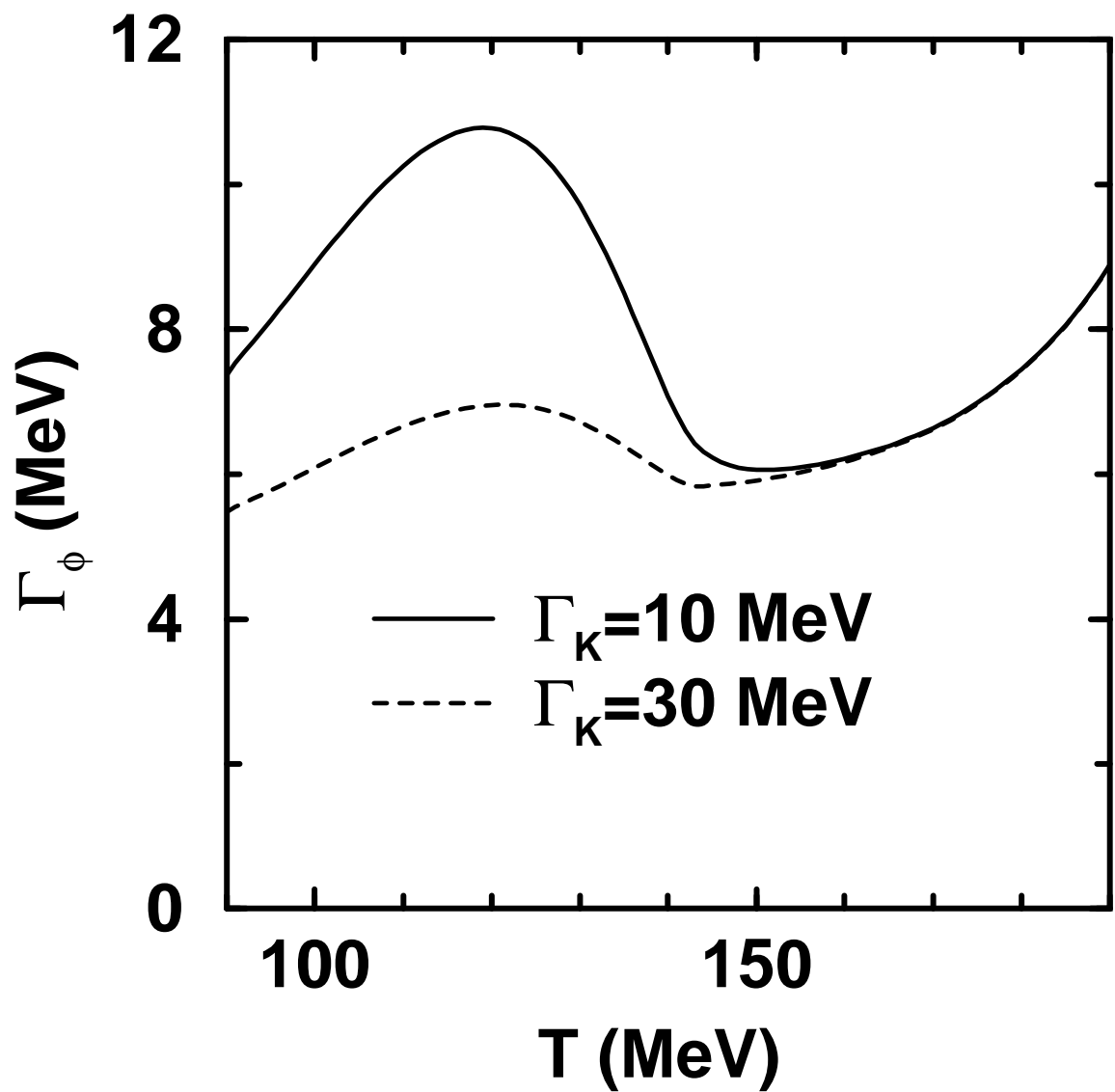
\end{document}